# Small angle neutron scattering contrast variation reveals heterogeneities of interactions in protein gels

A. Banc[1], C. Charbonneau[1], M. Dahesh[1,2], M-S Appavou[3], Z. Fu[3], M-H. Morel[2], L. Ramos[1]

[1] *Laboratoire Charles Coulomb (L2C), UMR 5221 CNRS-Université de Montpellier, F-34095 Montpellier, France*

[2] *UMR IATE, UM-CIRAD-INRA-SupAgro, 2 pl Pierre Viala, 34070 Montpellier, France.*

[3] *Jülich Centre for Neutron Science JCNS, Forschungszentrum Jülich, Outstation at MLZ, D-85747 Garching, Germany*

## Abstract

The structure of model gluten protein gels prepared in ethanol/water is investigated by small angle X-ray (SAXS) and neutrons (SANS) scattering. We show that gluten gels display radically different SAXS and SANS profiles when the solvent is (at least partially) deuterated. The detailed analysis of the SANS signal as a function of the solvent deuteration demonstrates heterogeneities of sample deuteration at different length scales. The progressive exchange between the protons (H) of the proteins and the deuteriums (D) of the solvent is inhomogeneous and 60 nm large zones that are enriched in H are evidenced. In addition, at low protein concentration, in the sol state, solvent deuteration induces a liquid/liquid phase separation. Complementary biochemical and structure analyses show that the denser protein phase is more protonated and specifically enriched in glutenin, the polymeric fraction of gluten proteins. These findings suggest that the presence of H-rich zones in gluten gels would arise from the preferential interaction of glutenin polymers through a tight network of non-exchangeable intermolecular hydrogen bonds.

# 1. Introduction

Small-angle scattering techniques are regularly used to probe the structure of polymers, colloids, surfactants and proteins dispersed in a solvent[1]. X-rays enable fast and localized measurements thanks to a high flux and a small beam size, but with a risk of radiation damage that is especially crucial for proteins. Radiation damage is however prevented with neutron scattering. In that case, deuterated solvents are usually used to increase the contrast between the solvent and the suspended objects. In principle, the contrast is due to differences in the scattering density of the suspended objects and the solvent, and small-angle X-ray and neutron scattering (SAXS and SANS) profiles are expected to be similar and to convey the same structural information. In addition, for multicomponent systems, contrast variation[1] permits for instance to selectively extinguish the signal of one of the component allowing one to selectively probe specific objects in multicomponent samples, e.g. the shell or the core of core-shell particles[2], the nanoparticles and polymers in nanocomposites gels[3], or the network heterogeneities in natural rubber[4]. Contrast variation has also been proven as a powerful technique to study biological structures in dilute regime[5], including protein-protein, protein-ribosome, and protein-DNA complexes[5, 6] or larger complexes as caseins[7]. However some precautions must be taken with the use of deuterated solvents since H/D exchange between the labile protons of the scattering objects and the deuterated solvent occurs[8]. This exchange, which might entail non uniform labeling in the case of proteins[9], has to be taken into account to quantitatively interpret small-angle neutron scattering data[8,10]. Deuteration of the solvent has been shown in particular to modify the temperature of phase transitions[11], to induce the clusterization of polymers in solution[12], and to modify the stability of proteins[29, 35, 36,13] due to a modification of the balance between intramolecular and hydration interactions[14]. Thus, when using deuterated solvent in a sample comprising proteins in order to enhance the contrast, one has to be aware of the possible alteration of the interactions at play and of the non-uniform labeling of the proteins, which might lead to misinterpretation of SANS data.

Here, we focus on gluten proteins extracted from wheat. Those proteins are among the most complex families of proteins due to their very broad polymorphism. They are mainly composed of 50% monomeric gliadins and 50% polymeric glutenins[15]. Gluten proteins are responsible for the remarkable viscoelastic properties of dough. However despite extensive studies in order to provide structural and mechanistic basis for the improvement of dough, there is still a crucial need to understand the supramolecular organization of gluten proteins and the link with the viscoelastic properties[16, 17]. Viscoelasticity is conventionally associated to disulfide bonds that form junction points between the proteins but several studies have also highlighted the important role played by non-covalent bonds as hydrogen bonds, hydrophobic and electrostatic interactions[18-20]. Most previous studies have been performed in water, while those proteins are not water soluble, rendering more complex the rationalization of the experimental results. Our approach is to use instead of water a solvent made of equal volumes of water and ethanol, a food-grade solvent in which a model extract of gluten proteins, composed of 45% gliadins and 55% glutenins, can be well dispersed and behaves as polymer chains in good solvent conditions[21]. At sufficiently high concentration, gels with remarkable viscoelastic properties are obtained[22].

In this paper, we investigate gluten proteins gels by scattering techniques. The starting point of this study is the finding that SAXS and SANS profiles markedly differ when using a (at least partially) deuterated solvent. Such mismatch between SAXS and SANS spectra has been rarely reported in the literature[23, 24] presumably due to the difficulty of rationalizing such findings. We provide here a consistent and quantitative rationalization of our experimental results gathered with various levels of solvent deuterations. Our analysis indicates an exchange between the protons H of the proteins and the deuteriums D of the solvent. The whole scattering data are interpreted by considering a heterogeneous H/D exchange due to localized zones where protein H/D exchanges would be prevented due hydrogen bonds. The size (60 nm) of those zones is comparable to that of the protein assemblies measured in the dilute regime, reflecting a characteristic size over which H/D exchange can be prevented[21]. Our conclusions are corroborated by spectroscopy and chromatography analyses performed on more dilute samples.

## 2. Materials and Methods

### 2.1. Materials

Native gluten powder (81.94% protein, dry basis) was courtesy of Tereos-Syral (France). A protein fraction representative of gluten in composition (glutenin/gliadin ratio = 1.1), soluble in ethanol/water (50/50, v/v), was extracted according to a protocol previously published by us[21], and freeze-dried. The exact composition of the fraction is detailed in Supporting Information (SI). The two main components are polymeric glutenins, which are composed of polypeptidic chains of high-molecular weight glutenin subunits (HMW-GS), and/or low-molecular weight glutenin subunits (LMW-GS) linked together by disulfide bonds, and monomeric gliadins, of different types (ω-gliadins, α/β and γ-gliadins). We used water purified from a milliQ system with a nominal resistivity of 18.2 MOhm.cm. Ethanol was of analytical grade. Deuterated solvents were purchased from Eurisotop. The isotopic enrichment of deuterated solvents was ≥99%, ≥99%, ≥ 99.97% for ethanol D6 ($C_2D_5OD$), ethanol OD ($C_2H_5OD$) and water ($D_2O$) respectively.

The scattering length densities (SLD) of the various solvents are given in Table 1. The SLD of species $i$ were calculated using $\rho_i = \sum_j \frac{b_j}{v_i}$. Here, $v_i$ is the volume of the species $i$, and the sum is performed over all atoms of the species with $b_j$ the coherent scattering length of the $j$-th atom.

|  | $H_2O$ | $D_2O$ | $C_2H_5OH$ | $C_2H_5OD$ | $C_2D_5OD$ |
|---|---|---|---|---|---|
| $\rho$ ($10^{-6}$ Å$^{-2}$) | -0.56 | 6.4 | -0.35 | 5.17 | 6.16 |

**Table 1.** Scattering length densities of the solvents

The distribution of the SLD of the protein extract in protonated solvent was calculated according to Jacrot[25] protonated amino-acid SLD values considering the polypeptide composition of the gluten extract (the detailed characterization of the composition is given in SI). We show in figure 1 the SLD distribution of the main polypeptide components of the polymeric glutenin, namely HMW-GS and LMW-GS, of the different gliadin polypeptides, ω-gliadins, α/β, γ-gliadins, and of the albumin/globulin proteins (alb/glo). The mean SLD value of the gluten protein extract calculated from the contribution of each polypeptide component is $\bar{\rho}_{\text{prot}} = (2.0 \pm 0.1)\ 10^{-6} \text{Å}^{-2}$. Here, the standard deviation takes into account the standard deviation on the SLD of each class of polypeptide but also a 5% uncertainty on their specific contribution to the total protein content of the wheat gluten fraction. We show also in figure 1 (red data) the SLD distribution once all the labile H of the proteins (i.e. the H that are bonded to nitrogen, oxygen or sulfur atoms[25]) have been exchanged with D. We find that the distribution is wider for deuterated proteins than for hydrogenated proteins due in particular to the larger amount of residues that can exchange more than one proton (e.g. arginine, tyrosine, threonine…) in HMW-GS. For deuterated proteins we find $\bar{\rho}_{\text{prot}}^D = (3.4 \pm 0.3)\ 10^{-6} \text{Å}^{-2}$.

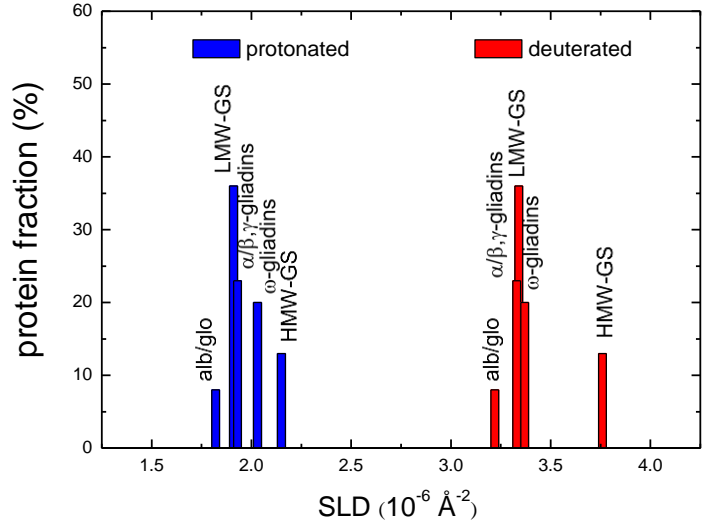

**Figure 1.** Distribution of the scattering length densities of the different classes of proteins present in the protein extract (as characterized in SI). Results are shown for fully hydrogenated proteins (blue) and for deuterated proteins (red) where all labile H have been exchanged with D.

**2.2. Sample preparation and composition**

Samples were prepared by dispersing the freeze-dried protein in an ethanol/water (50/50 v/v) solvent. For most samples, we used 290 mg of protein and 1ml of solvent. Considering a protein density of 1.32, the protein volume fraction of these samples was $\Phi=0.18$. Additional samples with $\Phi=0.04$ were also prepared. Samples with $\Phi=0.18$ were gel-like, whereas samples with $\Phi=0.04$ were fluid-like. The homogenization of all samples was performed on a rotary shaker overnight at room temperature. The samples were then stored at 20°C.

Various solvents were used. Water was a mixture of $H_2O$ and $D_2O$ with various molar fractions of $D_2O$, $x_{D2O}$. Similarly ethanol was either a mixture of $C_2H_5OH$ and $C_2D_5OD$ (D6 group samples), or a mixture of $C_2H_5OH$ and $C_2H_5OD$ (OD group samples), with various molar fractions of $C_2D_5OD$, $x_{C2D5OD}$, or $C_2H_5OD$, $x_{C2H5OD}$. The ethanol/water compositions of the samples investigated are summarized in figure 2. Samples are grouped in series. All series except Series I' were prepared with fully deuterated ethanol, $C_2D_5OD$. Series I and I' correspond to samples with an equal molar content of deuterated water and deuterated ethanol, but which comprise various deuterations. We define the deuteration as $(x_{C2D5OD} + x_{D2O})/2$, for D6 samples, and $(x_{C2H5OD} + x_{D2O})/2$, for OD samples. Samples in Series II, III and IV are characterized by a constant deuteration (50%, 63% and 75% respectively), but various molar fractions of $D_2O$ and $C_2D_5OD$.

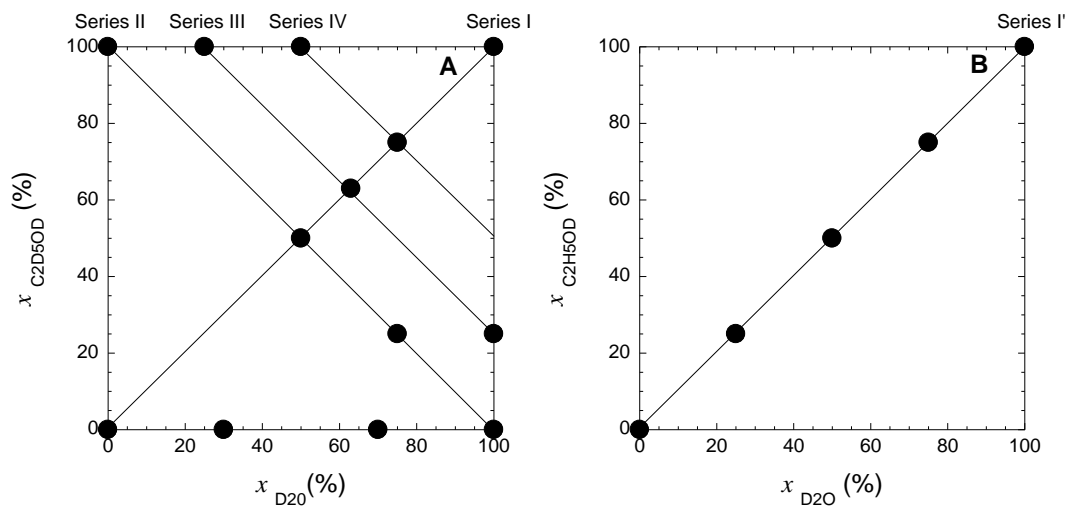

**Figure 2.** Water/ethanol (v/v 50/50) composition of the gluten gels studied by neutron scattering. (A) D6 group: samples prepared with a mixture of $H_2O/D_2O/C_2H_5OH/C_2D_5OD$ (B) OD group: samples prepared with a mixture of $H_2O/D_2O/C_2H_5OH/C_2H_5OD$.

## 2.3. Small-angle X-ray scattering

Synchrotron small-angle X-ray scattering (SAXS) experiments were conducted at Soleil, Saclay, France, on the Swing beam line. The samples were held in capillaries of internal diameter 1.5 mm. The beam energy was 12 keV and two sample-to-detector distances (1.5 and 5.5 m) were used, yielding scattering wave-vectors in the range ($1.2\ 10^{-3}$ - $7\ 10^{-1}$) $Å^{-1}$. The scattered intensity, $I(q)$, was obtained by using standard procedures, including subtractions of empty cell, solvent and background.

## 2.4. Small-angle neutron scattering

Experiments were performed on two instruments operated by JCNS at the Heinz Maier-Leibnitz Zentrum (MLZ, Garching Germany): KWS1 and KWS3. Small-angle neutron scattering (SANS) experiments were performed on KWS1[26] using three configurations with various wavelength, $\lambda$, and sample-detector distances, $D$, ($D$ = 20 m, $\lambda$ = 10 Å; $D$ = 8 m, $\lambda$ = 8 Å; and $D$ = 2 m, $\lambda$ = 8 Å) covering a $q$-range from $10^{-3}$ to 0.25 $Å^{-1}$. Very small-angle angle neutron scattering experiments running on the focusing mirror principle[27] were performed on the KWS3 instrument. Two sample-to-detector distances (1.2 m and 9.5 m) were used with a wavelength $\lambda= 12.8$ Å to access *q*-vectors from $2 \cdot 10^{-4}$ to $10^{-2}$ Å$^{-1}$. The samples were held in 1 mm-thick quartz cells. The reduction of raw data was performed by the routine qtiKWS[28] including corrections for detector sensitivity, background noise and empty cell signal. Absolute determination of scattering cross sections *I(q)* per unit sample volume in cm$^{-1}$ was obtained thanks to a calibration with a 1.5 mm-thick polymethylmethacrylate sample. Incoherent background was estimated using a far-point method and a linear evolution of incoherent background with sample deuteration was obtained.

## 2.5. Attenuated total reflectance - Fourier transform infrared spectroscopy

To evaluate the protein and solvent deuteration, experiments were performed on an Alpha Fourier transform infrared Bruker apparatus equipped with the single reflection diamond Attenuated Total Reflection (ATR) module. Spectra were recorded by the co-addition of 24 scans at a resolution of 8 cm$^{-1}$. For the analysis of proteins, samples were freeze-dried and maintained in an inert atmosphere in order to avoid contamination with the hydrogenated atmospheric water. $N_2$ was used to break the vacuum after freeze-drying and the spectrometer was placed in a glove box saturated in $N_2$ to avoid hydration of samples with the air humidity during measurements.

## 2.6. Size exclusion-high performance liquid chromatography

Protein size distribution was measured using size exclusion-high performance liquid chromatography (SE-HPLC) performed on an Alliance system equipped with a TSK G4000 SWXL column. Samples were diluted (at about 1mg/ml) in the elution buffer composed of 0.1 M sodium phosphate buffer at pH 6.8, 0.1% sodium dodecyl sulfate (SDS) and urea at a concentration of 6M. Elution of the injected sample (20μl) was performed at 0.7 ml/min and the detection of the different species was recorded at a wavelength of 214 nm. The apparent molecular weight calibration of the column was obtained using a series of protein standards with molecular weight in the range 13 to 2 000 kDa according to[21].

All measurements were performed at room temperature.

# 3. Results

### 3.1. SAXS-SANS mismatch

We show in figure 3 the scattering profiles of gluten gels samples prepared in three different ethanol/water solvents comprising various amounts of hydrogenated and deuterated compounds, namely a purely hydrogenated solvent ($x_{D2O} = x_{C2D5OD} = 0$), a purely deuterated solvent ($x_{D2O} = x_{C2D5OD} = 1$), and a mixture of heavy water and perdeuterated ethanol ($x_{D2O} = 1$, $x_{C2D5OD} = 0$). Surprisingly, the SANS profiles of the samples prepared with the three solvents markedly differ (figure 3A). Not only does the amplitude of the scattered intensity changes as the overall ratio of deuterium over hydrogen varies, as expected, but the shape of the scattering profile significantly varies as well. Interestingly, we find however that the SAXS profiles of the same three samples nearly overlap in the whole range of wave vectors investigated (figure 3B). In SAXS experiments, the contrast mainly arises from the differences in scattering length densities between the proteins and the solvent. The fact that all spectra nearly superimpose indicates that the spatial organization of proteins in the various solvents does not significantly differ. The SAXS scattering is characteristic of a polymer gel as previously described by us for the protein extract dispersed in a purely protonated solvent[21].

In the absence of deuterium in the sample, the contrast in SANS mainly arises from the contrast between protein and solvent (as expected from Table 1), and the SAXS and SANS profiles nearly superimpose in the whole range of wave-vector investigated (figure 3A) provided a normalization factor due to the different contrast probed in the two experiments is used. On the other hand, when the sample contains deuterium atoms, our results demonstrate that the SANS scattering is not only due to the contrast between the proteins chains and the solvent. We show below that a careful analysis of the scattered intensity for several series of samples with varying deuterium contents allows one to extract quantitative information on the interaction at play. Those interactions are hidden in SAXS data. In the following we uniquely focus on the SANS data and quantitatively analyze both the shape and the intensity of the scattering curves for all samples investigated.

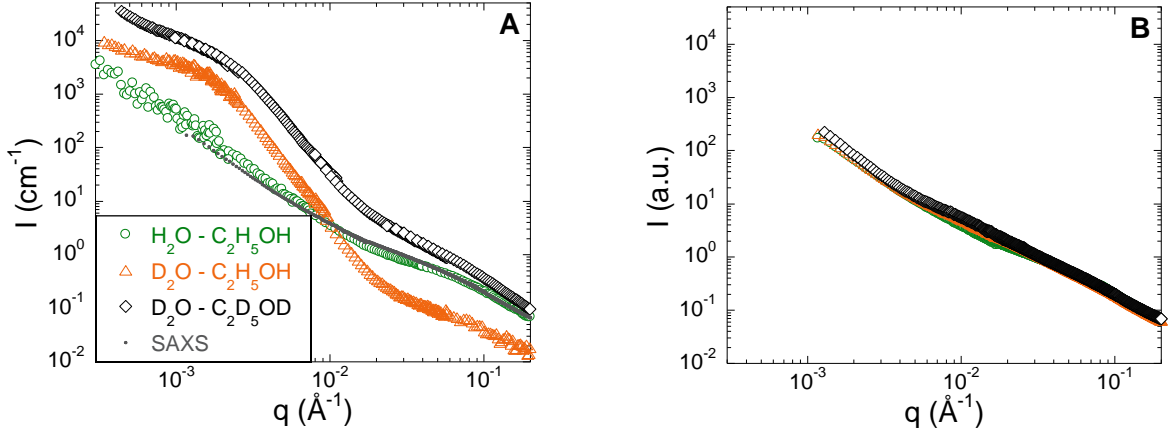

**Figure 3.** Small angle scattering spectra of gluten gels prepared with water-ethanol ($x_{D2O} = x_{C2D5OD} = 0$, green circles), heavy water-ethanol ($x_{D2O} = 1$, $x_{C2D5OD} = 0$, orange triangles) and heavy water-perdeuterated ethanol ($x_{D2O} = x_{C2D5OD} = 1$, black diamonds): (A) SANS spectra. The water-ethanol SAXS spectrum (in grey) is superimposed on the SANS spectra for comparison. (B) SAXS spectra. Symbols are the same as in (A).

## 3.2. Characteristic length scales

As shown in figure 3, the SANS spectra evolve dramatically as the solvent deuteration varies. Here, we investigate more quantitatively the effects of the overall deuteration of the solvent on the shape of the scattering curve. Figures 4A and 4B display the evolutions of the SANS spectra for samples from Series I (with $C_2D_5OD$ as deuterated alcohol) and Series I' (with $C_2H_5OD$ as deuterated alcohol). The labelling by the two kinds of deuterated ethanol molecules is different as $C_2H_5OD$ contain one labile deuterium that produces eventually protonated ethanol by H/D exchange whereas $C_2D_5OD$ contains, in addition to the labile deuterium bonded to oxygen, five deuterium atoms irreversibly linked to the carbon atoms that permanently label the ethanol molecules. In Figures 4A and 4B the deuteration of the solvent varies from 0 to 100%. In the case of a fully hydrogenated solvent, the shape of the scattering curves can be simply related to the protein structure: at large wave vector, the scattering of the random walk of individual polymer chains is measured, and the scattering intensity $I$ scales as $q^{-2}$ as expected and observed for denaturated or intrinsically disordered proteins[29]. A cross-over to a plateau regime for length scales larger than the blob-size is measured. Finally for very small length scales a power law evolution of the scattering is recovered, which has been ascribed to the fractal organization of the protein inhomogeneities in the gel. In accordance, the whole scattering curve can be very well fitted (figure 4) with the empirical functional form:

$$I(q) = \frac{A}{1+(q\xi)^2} + Bq^{-n} \quad \text{(Eq. 1)}$$

Here the first term on the right hand side that dominates at large $q$ is a standard Orstein-Zernike term. This term accounts for the concentration fluctuation of polymer inside a blob of size $\xi$, for a semi-dilute polymer solution in the regime $q\xi > 1$ [30, 31]. The second term that dominates at small $q$ accounts for the fractal organization of the protein at large length scale (with fractal dimension $n$). Best fits of the data yield $n=2$. Note that Eq. 1 has also been previously used to account for the scattering of methylcellulose solutions[32], POE solutions[31], PEG hydrogels[33], and peptide hydrogels[34].

From figure 4A and 4B the most evident evolution of the shape of the profile with solvent deuteration is the $q$-dependence of the scattering intensity at small wave vector ($q < \sim 0.01$ Å$^{-1}$), whereas the signal at small length scale seems similar in all cases. At small $q$, the scattered intensity strongly increases as $q$ decreases, with a power law with an exponent close to -4, and eventually reaches a plateau at even lower $q$. Accordingly, in the whole range of $q$, the data are fitted with the empirical functional form:

$$I(q) = \frac{A}{1+(q\xi)^2} + \frac{C}{[1+(q\Xi)^2]^2} \quad \text{(Eq. 2)}$$

Here the first term is similar to the one used above to account for the signal of polymer chains in a fully hydrogenated solvent (Eq. 1), whereas the second term is a Debye-Büeche term, originally used to describe inhomogeneous solids[35]. This second term is characterized by a correlation length $\Xi$, and a Porod behavior ($I(q) \sim q^{-4}$) at $q\Xi \gg 1$ that assumes smooth interfaces. Note that Eq. 2 has been previously used to describe various polymer materials including synthetic polymer gels[36, 37], gelatin gels[38], and natural rubber[4]. The scattering profile of the samples from Series I and I' with (at least partially) deuterated solvents can be very well fitted with Eq. 2 (figures 4A and 4B). This equation provides also very good fits of the scattering profiles of the other (at least partially) deuterated samples investigated (figures 4C, 4D, and 4E).

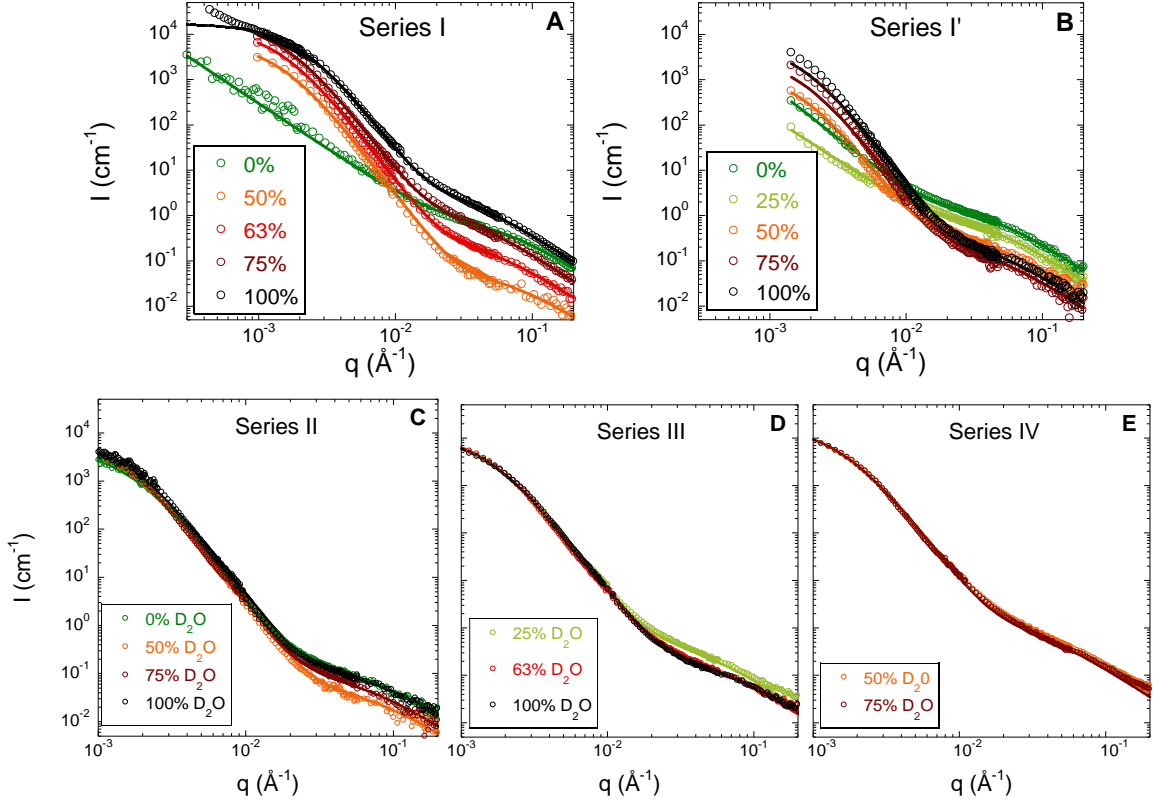

**Figure 4.** SANS spectra of gluten gels for different levels of solvent deuteration (A, Series I, and B, Series I') and different solvent mixtures with equal solvent deuteration (C, D, E). Symbols are experimental data points and lines are the best fits using Eq. 1 or Eq. 2.

The two characteristic length scales, $\xi$ and $\Xi$, extracted from the fits with Eq. 1 and Eq. 2 are gathered in figure 5, for samples of the D6 group (figure 2). We show in figure 5A the evolution of the large scale characteristic length, $\Xi$, which can only be measured in the presence of a deuterated solvent. We mention that not all samples were measured on a very broad range of wave-vectors. For some samples (Series I', figure 4B), data at small wave vectors are not available. Hence the plateau of the scattered intensity at small $q$ is hardly measured and reliable measurements of the characteristic size $\Xi$ are not accessed. When $\Xi$ can be evaluated, we find that $\Xi$ is roughly constant ($\Xi = (600 \pm 100)$ Å), independently on the samples investigated. Consequently, for the samples of Series I' the fit of the data using Eq. 2 are performed by imposing for $\Xi$ the average numerical value found experimentally for other samples. On the other hand, reliable measurements of the blob size $\xi$ are obtained for all samples. Figure 5B shows the evolution $\xi$ with the solvent deuteration for samples of the D6 group. We find that $\xi$ is constant ($\xi = (15 \pm 5)$ Å) for solvent deuteration up to 50% and steadily increases with solvent deuteration, reaching 30 Å for a fully deuterated solvent. The evolution of the blob size with solvent deuteration can be interpreted as resulting from an evolution of the protein flexibility. Indeed, for polymer chains in good solvent conditions, the scaling theory[30] predicts $\xi = l_0 \Phi^{-3/4}$, where $l_0$ is the polymer persistence length, or monomer size for a flexible polymer, and $\Phi$ is the volume fraction of polymer. Here $\Phi = 0.18$, yielding a persistence length that varies from 4 to 8 Å with solvent deuteration. These numerical values are in excellent agreement with our previous measurement for hydrogenated samples with various concentrations[21] and with the values experimentally found for unstructured proteins (between 5 and 7 Å)[39]. Note in addition that a stiffening of the protein chain with solvent deuteration has been measured by force spectroscopy for proteins similar to ours although simpler (model peptide of the repetitive domain of glutenins)[40]. Other studies have evidenced the influence of heavy water on the protein rigidity, with a rigidity that could increase [41,42] or decrease in the presence of $D_2O$[42] depending on the overall hydration of the proteins.

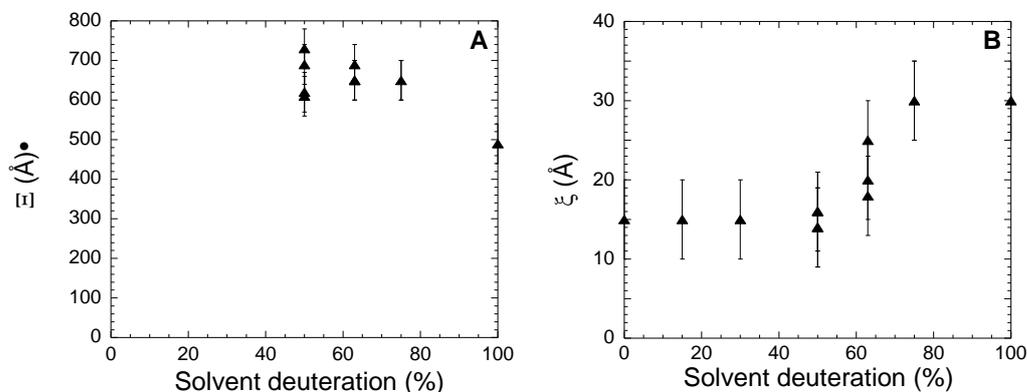

**Figure 5.** Large scale characteristic size $\Xi$ (A) and blob size $\xi$ (B), as a function of the solvent deuteration for samples of the D6 group.

### 3.3. Heterogeneous deuteration of the proteins

The distribution of scattering length densities of the protonated gluten protein extract is given in figure 1. When the solvent is partially or totally deuterated, some protons of the proteins are replaced by deuterium through exchange with the solvent. The potentially exchangeable hydrogens are the most labile ones, which are bonded to nitrogen, oxygen or sulfur atoms[25]. For those labile hydrogens, the extent of exchange depends on the H/D stoichiometry and the accessibility of the labile protons of the proteins for the deuterium of solvent. The protein deuteration in our samples is estimated by ATR-FTIR spectroscopy following the intensity of the amide II band that occurs at 1550 cm$^{-1}$ for protonated amide groups, and shifts down to 1450 cm$^{-1}$ for deuterated groups (amide II'). Figure 6A displays the spectra in the amide II region of gluten gel samples with different concentrations of deuterium. A clear decrease of the amide II bonds occurs as the concentration of deuterium in the sample increases, showing unambiguously the exchange of deuterium between the solvent and the hydrogen involved in the amide II bonds, as classically observed for proteins[43, 44]. In our experiments, there are three sources of deuterium $D_2O$, $C_2D_5OD$ and $C_2H_5OD$, and up to two different sources for one given sample, i.e. $D_2O$ and/or ($C_2D_5OD$ or $C_2H_5OD$). Only the labile deuteriums of the solvent could be exchanged with the hydrogens of proteins. The two deuteriums of the heavy water are labile, whereas only the deuterium linked to the oxygen of the ethanol molecule is labile. Interestingly, when the amide II absorbance is plotted as a function of the concentration of labile deuterium all the data acquired with the various solvents fall on a unique curve. We find that the amide II absorbance decreases linearly with the concentration of labile deuterium in the sample, demonstrating that the protein deuteration is proportional to the labile D content of the sample whatever the origin of deuterium.

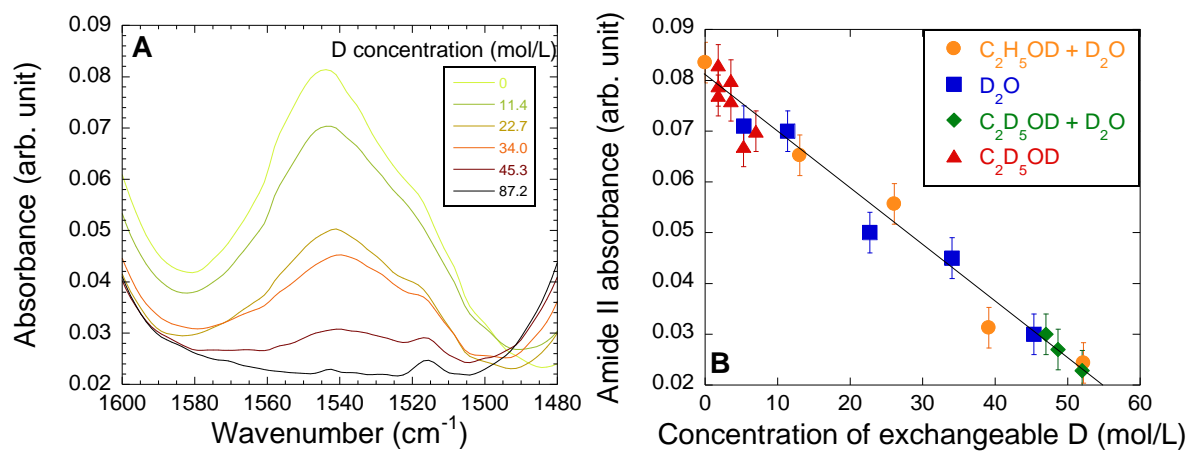

**Figure 6.** FTIR analysis of protein deuteration in gluten protein gels. (A) FTIR spectra in the amide II region for samples comprising different total concentrations of deuterium as indicated in the legend. The source of deuterium is $C_2D_5OD$ and/or $D_2O$. (B) Evolution of the amide II band intensity as a function of the concentration of labile deuterium for samples prepared with different water/ethanol (50/50) solvents. The legend indicates the nature of deuterated solvent used. Symbols are experimental points and the line is a linear fit of the data.

At large wave vectors, the SANS signal that probes the local structure of the samples is theoretically proportional to the contrast between proteins and solvent, $(\bar{\rho}_{\text{prot}} - \rho_{\text{solv}})^2$, where $\bar{\rho}_{\text{prot}}$ is the average SLD of the protein, and $\rho_{\text{solv}}$ is the solvent SLD. Varying the solvent scattering length density (SLD) should in principle allow one to extinguish the signal of the proteins once $\rho_{\text{solv}}$ equals $\bar{\rho}_{\text{prot}}$, as in a standard contrast variation procedure. However the H/D exchanges between protons from polypeptides and deuteriums from the solvent have to be considered and both polypeptides and solvent SLD cannot be calculated a priori. As a consequence, we plotted in figure 7 the scattered intensity at large wave vectors (for convenience we report the intensity measured at $q=10^{-1}$ Å$^{-1}$) as a function of the average sample SLD ($\bar{\rho}_{\text{sample}}$), for samples of group D6 (figure 7A) and of group OD (figure 7B). Note that the range of sample SLD is much narrower for samples of OD group than for sample of D6 group due to the difference in the total deuterium content for the two ethanol used. In both cases, the curve displays a minimum, as expected from a contrast variation procedure, which appears at the average sample SLD $\rho^0 = (2.7\pm0.1)\ 10^{-6}$ Å$^{-2}$ for the D6 group, and $\rho^0 = (2.6\pm0.1)\ 10^{-6}$ Å$^{-2}$ for the OD group. This minimum corresponds to the best matching of the SLD of the proteins by that of the solvent, and at this point: $\rho_0 = \bar{\rho}_{\text{sample}} = \rho_{\text{solv}} = \bar{\rho}_{\text{prot}}$. Interestingly we observe that this minimum is different to zero in both cases suggesting a non-uniform SLD for all the polypeptides present in the sample[25]. Accordingly, the data of the scattering intensity at large $q$ are fitted with the functional form:

$$I(q=10^{-1}\text{Å}^{-1}) = K\ (\bar{\rho}_{\text{prot}} - \rho_{\text{solv}})^2 + I^0 \quad \text{(Eq. 3)}$$

Here $K$ is a constant related to the osmotic modulus of the sample and $I^0$ is the non-zero intensity at the matching point related to the distribution of protein SLD.

In order to fit the data with $K$ and $I^0$ as fitting parameters, we have to calculate the evolution of the solvent and the proteins SLD as a function of sample deuteration (figures 7C and 7D). Because of the exchange between deuterium and hydrogen, the scattering length density of the proteins, $\bar{\rho}_{\text{prot}}$, changes with that of the solvent. We expect $\bar{\rho}_{\text{prot}}$ to increase as the solvent deuteration increases. Infrared spectroscopy shows a linear decrease of the amide II band with the concentration exchangeable deuterium brought by the solvent (figure 6B). It is therefore reasonable to assume that $\bar{\rho}_{\text{prot}}$ increases linearly with the concentration of exchangeable deuterium in the sample. This concentration is proportional to the solvent SLD in the case of sample from group OD, as all the deuteriums brought by C$_2$H$_5$OD and by D$_2$O are labile. Hence for samples from group OD, $\bar{\rho}_{\text{prot}}$ is expected to vary linearly with the solvent deuteration. Knowing the value of protein SLD in a purely hydrogenated sample ($\bar{\rho}_{prot}$ = 2.0 $10^{-6}$ Å$^{-2}$) and the value at the matching point ($\bar{\rho}_{prot} = \rho_{solv} = \rho_{sample} = \rho^0$ =(2.6±0.1) $10^{-6}$ Å$^{-2}$), the linear evolution of $\bar{\rho}_{prot}$ and $\rho_{solv}$ as function of the solvent deuteration can be computed (figure 7D). The solvent SLD in the sample is 5% lower than the value calculated without exchange, while the average protein SLD increases from (2.0±0.1) $10^{-6}$ Å$^{-2}$ in the protonated solvent to (2.8±0.1) $10^{-6}$ Å$^{-2}$ in the fully deuterated solvent. The evaluation of $\bar{\rho}_{prot}$ with solvent deuteration is not as straightforward for samples of group D6 as all the deuteriums brought by the solvent are not exchangeable (one out of the 6 deuteriums of ethanol $C_2D_5OD$ can be exchanged). The evolution of $\bar{\rho}_{prot}$ is calculated by computing for each solvent deuteration the number of exchangeable deuterium and using the linear relation found in the case of samples of group OD. The results (figure 7C) show a nonlinear evolution of $\bar{\rho}_{prot}$ with the solvent deuteration. Because of this nonlinear evolution, the contrast does not vary symmetrically on each side of the matching point, and hence one does not expect a parabolic evolution of the scattered intensity at large $q$ with the solvent SLD, as observed experimentally (figure 7A). For fully deuterated solvents, we find that the SLD of the protein is (2.8±0.1) $10^{-6}$ Å$^{-2}$. This value is smaller than the expected theoretical values for the SLD of proteins for which all the labile hydrogen have been exchanged with deuterium (3.4 ± 0.3) $10^{-6}$Å$^{-2}$ (cf. section 2.1). Our data suggest that (57±7) % of the labile H of the proteins have been exchanged. This value is in the range of the H/D exchange level measured for other proteins by mass spectroscopy[45].

Finally, using the SLD values of figure 7B and C, Equation 3 provides nice fits of the experimental data (figure 7A and 7B), with comparable fitting parameters for the two sets of data, showing the consistency of our interpretation. We find $I^0$=(0.026 ±0.001) cm$^{-1}$ and $K$=(3.1±0.1) $10^{-22}$ cm$^3$, for samples of series I and $I^0$=(0.033±0.001) cm$^{-1}$ and $K$=(3.0±0.1) $10^{-22}$ cm$^3$, for samples of series I'.

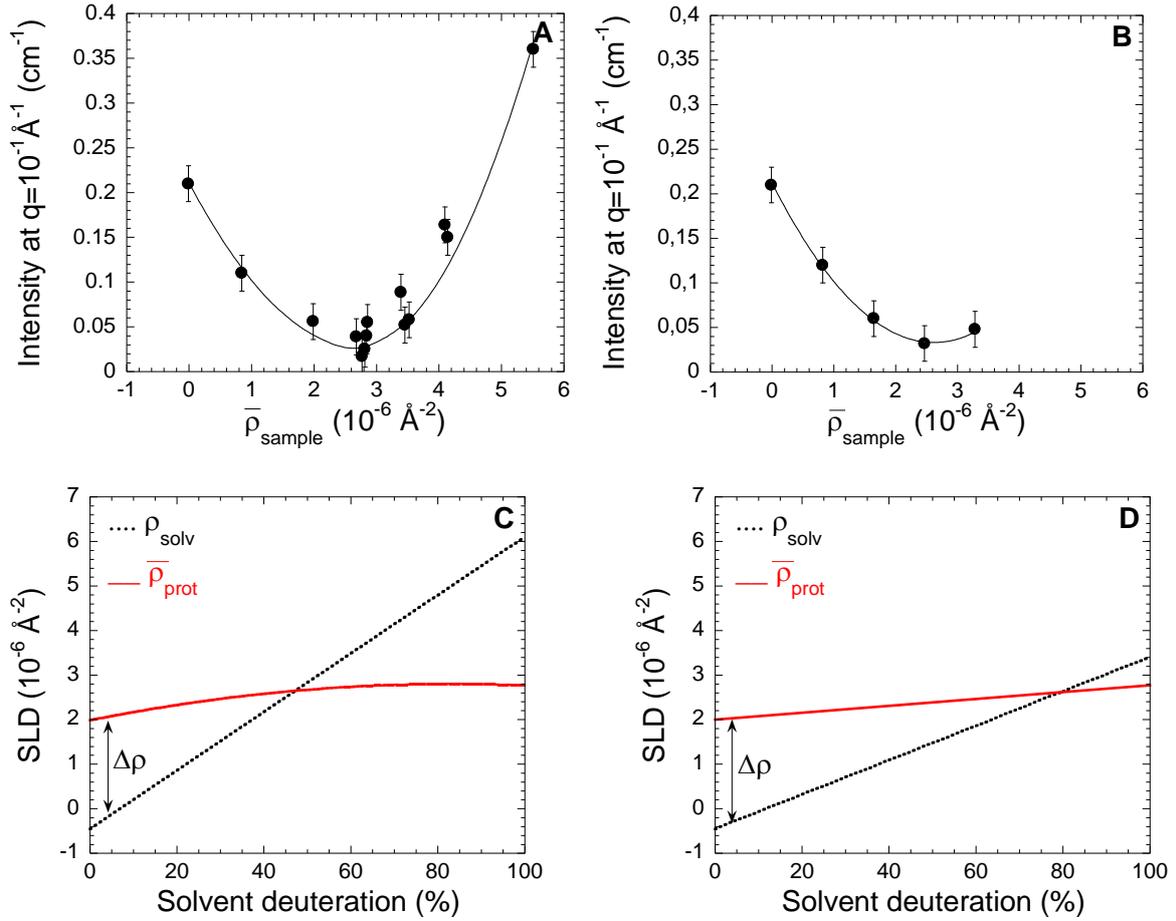

**Figure 7.** Contrast variations in the high $q$ regime. (A, B) Evolution of the SANS intensity at $q=10^{-1}$Å, as a function of the solvent SLD for samples prepared with a mixture of (A) $H_2O/D_2O/C_2H_5OH/C_2D_5OD$ (group D6), and (B) $H_2O/D_2O/C_2H_5OH/C_2H_5OD$ (group OD). The symbols are experimental data points and the lines are fits with Eq.3 using the SLD values shown in (C) for group D6 and in (D) for group OD. (C, D) Evolution of the solvent and protein SLD with the solvent deuteration as deduced from the analysis of FTIR results and from the solvent composition for samples of group D6 (C) and of group OD (D).

Using the same arguments, one can explain the evolution of the amplitude of the scattering at large $q$ for samples from Series II, III and IV (figures 4C, 4D, 4E). In these series the overall deuterium content is kept constant while the origin of deuterium differs. Because water molecules contain 2 times more exchangeable deuteriums than ethanol molecules, the amount of exchangeable deuterium varies along a series. Hence one expects a change of the SLD of the proteins, and consequently of the contrast $(\bar{\rho}_{prot} - \rho_{solv})^2$. We show in figure 8A, the evolution of $\rho_{solv}$ and $\bar{\rho}_{prot}$ (calculated as explained above) with the concentration of exchangeable deuterium in the sample, and we show in figure 8B the evolutions of the intensity at $q=10^{-1}$ Å$^{-1}$ as a function of the concentration of labile deuteriums in the samples. For samples of Series III and Series IV, $\rho_{\text{solv}} > \bar{\rho}_{\text{prot}}$. Hence as the amount of labile deuterium increases, the contrast decreases, as observed experimentally. By contrast, for samples of Series II, $\rho_{\text{solv}}$ and $\bar{\rho}_{\text{prot}}$ are expected to cross for a concentration of exchangeable D of about 45 mol/L. Consequently, one expects the contrast to vary in a non-monotonic fashion with the concentration of labile D, in full agreement with our experimental observations, although our data suggest a minimum contrast at a slightly lower concentration (~ 30 mol/L). On the other hand, comparable and non-zero values ($I^0 \approx 0.03$ cm$^{-1}$) are measured for the minimum contrast expected to be reached when $\bar{\rho}_{\text{prot}} = \rho_{\text{solv}}$.

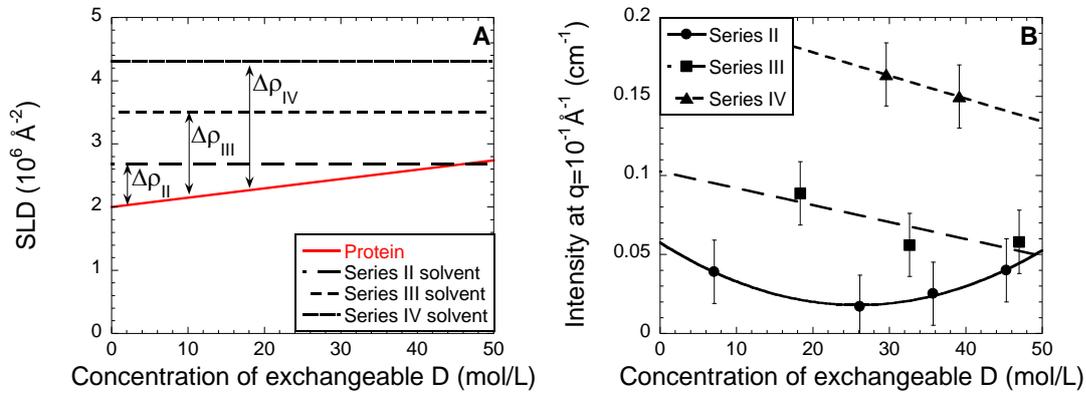

**Figure 8.** Intensities at $q=10^{-1}$ Å$^{-1}$ (D), and evolutions of the solvent and protein SLD (E) as a function of the concentration of exchangeable deuterium. Lines in (B) are guides for the eyes.

As mentioned above, a non-vanishing contrast implies a non-uniform SLD of the protein in the samples. A first naïve interpretation would be that the non-zero contrast originates from the polymorphism of the proteins comprising our protein extract. Considering a distribution of $N$ proteins characterized by the fractions, $f_i$, and the SLD, $\rho_i^D$ of the $i$ proteins in the deuterated solvent, $I^0$ can be estimated as [8] $S = K \sum_{i=1}^{N} f_i (\rho_i^D - \rho^0)^2$, with $\rho^0$ the matching SLD value, $\rho^0 = 2.6 \cdot 10^{-6}$ Å$^{-2}$. Using the distributions of protein SLD evaluated in section 2.1, one evaluates that $S$ is about 100 times lower than the experimental value of $I^0$, excluding the distribution of the protein SLD in the samples as unique origin for our experimental results. Instead, the concomitance of protonated and deuterated proteins in the partially deuterated solvent has to be taken into account to interpret the data. This suggests an inhomogeneous deuteration of proteins in ethanol/water solvents.

## 3.4. Large scale H-rich zones in a deuterated solvent

The scattering signal at low wave vectors ($q < 2 \cdot 10^{-2}$ Å$^{-1}$) evolves in terms of shape and intensity with solvent deuteration whereas it remains constant for a given level of deuteration of the sample (figure 4). Sample from Series II, III and IV were prepared with D6 ethanol, which contain deuterium atoms permanently bonded along the aliphatic chain, ensuring a high contrast between ethanol and water molecules after H/D exchange. Varying the origin of deuterium for a constant level of deuteration (samples from Series II, III and IV), an evolution of the SANS signal would be expected with the formation of ethanol-rich zones, in contrast with our experimental observations. Consequently, the invariance of the SANS spectra shows that the low $q$ signal cannot be attributed to protein induced ethanol-water demixion. Instead, the solvent presumably forms a homogenous mixture through the samples, at least at the length scales investigated in our experiments. Those length scales are indeed too large to probe the solvent clusters that may form in water/ethanol mixtures[46-48].

As shown in section 3.2 and in figure 4, for partially deuterated samples (from 50% deuteration), the low-$q$ regime of the SANS spectra is correctly fitted with the Debye-Bueche formalism that is generally used to account for the scattering of two-phase systems. The prefactor $C$ (eq. 2) is related to the characteristic parameters of the two phases:

$$C = 8\pi \Xi^3 (\rho_1 - \rho_2)^2 \varphi(1-\varphi) \quad \text{(Eq. 5)}$$

Here $\rho_1$ and $\rho_2$ are the SLD of the two phases, and $\varphi$ and $(1-\varphi)$ their volume fractions. Figure 9 displays the evolutions of the parameter $C/\Xi^3$ with sample deuteration. The large increase of $C/\Xi^3$ and its parabolic evolution with the solvent deuteration suggest that the contrast, $(\rho_1 - \rho_2)$, is proportional to the sample deuteration, while $\varphi$ is constant. The increase of the contrast between the two phases suggests that, upon increasing solvent deuteration, one phase is more easily deuterated than the other one. The hypothesis of contrasted phases due to different protein concentration and/or spatial organization can be discarded. Indeed, those configurations would imply an electronic density contrast between the two phases that would lead to differences in low q for the SAXS profiles of samples prepared with different deuteration of the solvent, in contrast with our experimental observations (fig. 3B). Hence, the two phases should correspond to zones with equivalent protein concentration and organization but with different deuteration levels, one of the phases being more hydrogenated at the expense of the other one. Unfortunately, our data cannot allow the independent evaluations of the SLD of the two phases ($\rho_1$ and $\rho_2$) of their respective volume fractions ($\varphi$ and $(1-\varphi)$).

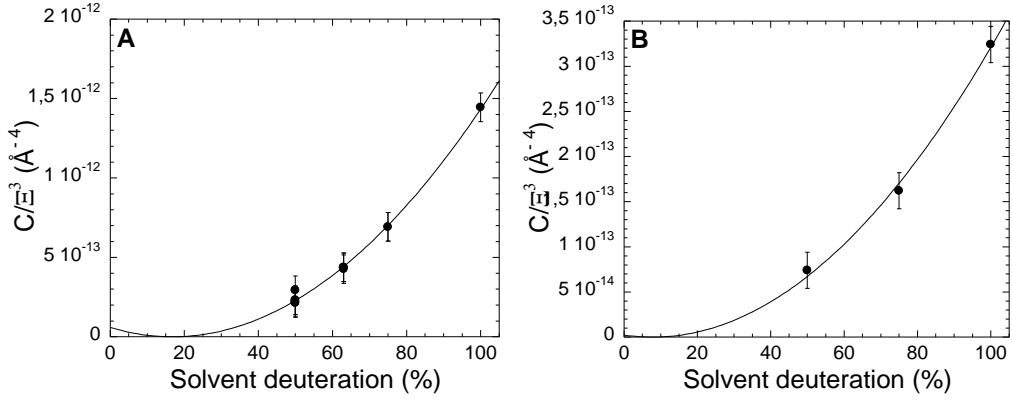

**Figure 9.** Evolution of ($C/\Xi^3$) with the solvent deuteration, for samples of the D6 group (A) ad of the OD group (B). Symbols are data points and the continuous lines are parabolic fits.

To validate the existence of H-rich zones, complementary experiments have been performed with samples prepared at lower protein volume fraction ($\Phi$=0.04) with the aim to isolate and characterize such zones. In a fully protonated solvent the low concentration sample is homogeneous while phase-separation in two phases is observed in per-deuterated solvent (pictures are given in SI). The volume of the upper phase is about 4 times larger than the volume of the bottom phase, and protein volume fractions, as determined using chromatography, are $\Phi$=0.036 in the upper phase, and 0.054 in the bottom phase, discarding the hypothesis of protein aggregation.

To evaluate and compare the amount of deuteration in the two phases, ATR-FTIR spectroscopy was performed. FTIR spectra (figure 10A) display the strong stretching bands of the deuterated solvent ($\nu_{OD}$, $\nu_{CD_2}$, $\nu_{CD_3}$) and amide bands of proteins. A tiny OH stretching band, attributed to the protonated proteins that potentially undergo H/D exchange is also observed. OH, respectively OD, stretching bands, measured in the (3100-3600) cm$^{-1}$, respectively (2100-2650) cm$^{-1}$, ranges are used to define the ratio $r$ as: $r = \dfrac{\left[\dfrac{A_{OD}}{A_{OH}}\right]_{top}}{\left[\dfrac{A_{OD}}{A_{OH}}\right]_{bot}}$

Here $A_i$ is the integration area of the band $i$, ($i$=OD or OH), and the subscript "*top*" and "*bot*" design the top and bottom phases in the two-phases sample. A quantitative analysis of the FTIR spectra of the two phases yields $r$=1.5±0.3, indicating that the top phase is more deuterated than the bottom phase. To determine the implication of the proteins, ATR-FTIR is also performed on the freeze dried proteins extracted from each phase. An inert atmosphere is here used to freeze-dry the proteins and perform spectroscopy in order to avoid hydration and protonation of the dried proteins by the air humidity. Normalized infrared spectra (figure 10B) reveal a higher intensity of amide II and OH stretching bands for the proteins extracted from the bottom phase than for the proteins extracted from the top phase. Measurements of several samples confirm that these differences of intensity although weak are significant. In addition, a quantitative analysis gives $r=2.0\pm0.3$, a numerical value in quantitative agreement with the one found above. Hence, the FTIR results demonstrate that proteins are more deuterated in the top phase than in the bottom phase. Concerning the amide I band, our results do not evidence any significant difference between the spectra of the two phases whereas a very small shift towards low wavenumbers is observed for proteins from the bottom phase, suggesting a few more β-sheet secondary structures. In addition, the OH and OD stretching bands are shifted to low wavenumbers in the spectra of the freeze dried proteins compared to the spectra of the protein suspensions, indicating bonded hydroxyl groups through hydrogen-bonds.

Finally, we also note that the total quantity of hydroxyl groups in the protein spectra seems more important for the proteins of the bottom phase than for those of the upper phase. This could be attributed to the different content of residues with hydroxyl side groups (Tyrosine, Threonine, Serine, Glutamic acid) in the various polypeptides from gluten, and/or would suggest that more solvent molecules are involved in the first hydration shell of proteins from the bottom phase. More experiments would however be required to confirm this statement.

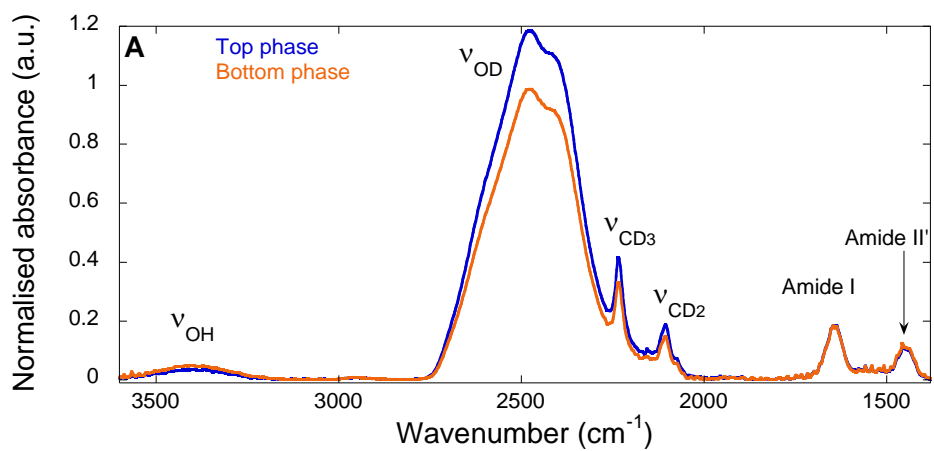

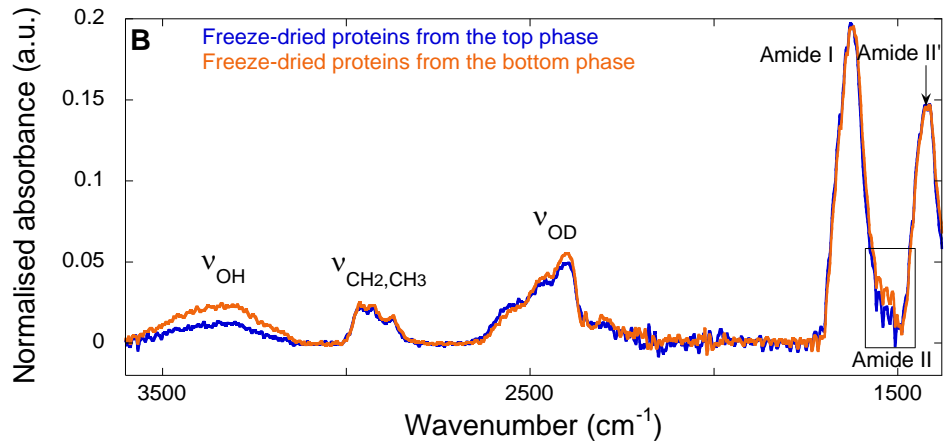

**Figure 10.** FTIR spectra of the two phases from the dilute D sample: top (blue line) and bottom (orange line) phases (A) spectra of the two phases (B) spectra of the freeze-dried proteins extracted from the two phases. Intensities are normalized by the amide I band.

Finally, as the protein extract is composed of gliadins and glutenin polymers, the two main subclasses of gluten proteins, the protein composition in each phase was also studied by chromatography. The HPLC analysis of phases reveals that the bottom phase is enriched in glutenin polymers, compared to the upper phase which is richer in gliadins (figure 11). The mass ratio of glutenin over gliadin is 1.1 in the top phase and 1.7 in the bottom phase. This suggests an important role of the protein polymorphism on the heterogeneous deuteration of the protein gel.

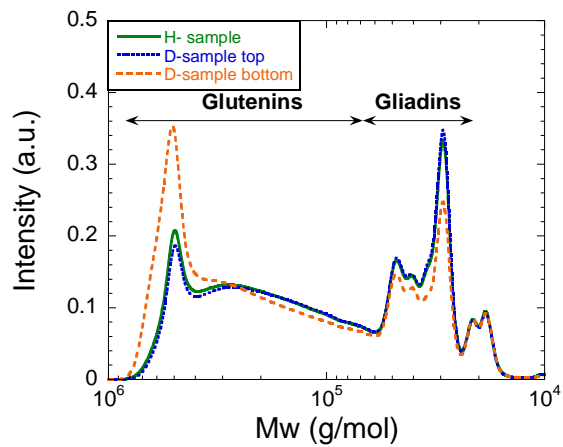

**Figure 11.** Pictures and HPLC profiles of samples prepared at $\Phi=0.04$ with different solvents: $H_2O/C_2H_5OH$ (50/50 v/v) for the H-sample, $D_2O/C_2H_5OD$ for the D-sample.

# 4. Discussion

We have shown that changing the level of deuteration of the solvent of gluten gels leads to drastic modifications of the shape and intensity of SANS spectra, while SAXS spectra remain unchanged. These findings indicate contrast effects rather than an evolution of the structure of the protein sample with solvent deuteration. For gels prepared with protonated solvents, the contrast mainly arise from the difference in scattering length density between the protein and the solvent, and SAXS and SANS spectra can be accounted for by the scattering of a polymeric gel characterized by a blob size and a fractal dimension of 2 at large length scales. When the solvent is deuterated, the SANS scattering profiles indicate a contrast between two phases with sharp interfaces and a characteristic length scale $\Xi$ of the order of 600 Å. The contrast analysis of the SANS signal and the comparison between the SAXS and SAXS spectra reveal that the two phases differ essentially by their contrast, one being more protonated than the other. This result is consistent with a heterogeneous deuteration of the gluten protein components evidenced by the imperfect matching of the protein in the high $q$ regime of the neutron scattering profiles. In full agreement with those observations, the spectroscopic analysis of the two co-existing phases obtained at a lower concentration in deuterated solvent reveals that proteins are more protonated in the minor bottom phase. The phase separation induced by the solvent deuteration can be tentatively interpreted in term of an unbalance of hydrogen bonds, as follows. In a fully hydrogenated solvent, the hydrogen bonds between the solvent and the proteins, and those between proteins, are assumed to be balanced, since proteins have been shown to behave as polymers in good solvent conditions. In deuterated solvents, , hydrogen bonds between the proteins and the deuterated solvent might be depleted at the benefit of hydrogen bonds between hydrogenated proteins, and between proteins and protonated solvent molecules. This unbalance could lead to a phase separation between a phase enriched in protonated proteins (forming hydrogen bonds with protons) and a phase comprising more deuterated proteins. We propose that the same process occur in more concentrated samples but that macroscopic phase separation is hindered by protein gelation. Hence, the protein domains involved in intermolecular H-bonds would be likely less available to undergo H/D exchange with the solvent and would contribute to delineate the large scale H-rich domains (600 Å), which are probed by SANS. Interestingly the size of these domains is similar to the size of protein assemblies measured in the dilute regime in a fully hydrogenated solvent[21]. Another remarkable feature of the phase-separated dilute sample is that the two phases display different protein compositions: the minor H-rich phase contains more glutenins. Hence a partitioning of both families of proteins (gliadins and glutenins) is evidenced whereas no partitioning of ethanol/water could be detected. This suggests different interaction parameters between the different gluten polypeptides, and between proteins and the different H/D solvents. We note that the liquid-liquid phase separation of a gluten protein extract in protonated ethanol/water was previously analyzed by Boire et al.[49] The authors showed that the partition of the proteins between the two phases depends on the molecular weight of the glutenin polymers, as for neutral polymers, whereas the different gliadins display the same interaction parameter. We suggest that this partition is mediated by the proteins through their interaction with the solvent. Gluten proteins contain indeed many glutamine residues that are prone to form intramolecular hydrogen bonds through β-turns and intermolecular hydrogen bonds with solvent molecules and with other polypeptidic chains[43] through intermolecular β-sheets[50]. These residues are concentrated in the repetitive domains of the polypeptide sequence of glutenin subunits in the form of hexapeptide and nonapeptides repeats[51]. On the basis of this feature and from the spectroscopic analysis of secondary structures of gluten proteins as well as from their H-NMR relaxation time upon hydration, Belton[18] developed a model of gluten elasticity. This model considers non-bonded mobile domains along the glutenin polymers where residues are well hydrated by the solvent and bonded regions <u>relevant</u> for interchain hydrogen bonds Interestingly, a significant difference in the dynamical behavior in protonated and deuterated solvent was previously observed for C-hordeins, the barley homologous of ω-gliadins with a large repetitive sequence rich in glutamines[52]. The modification of solvent-protein interactions in deuterated solvent via hydrogen bonds of glutamines could be also at the origin of the phenomena observed.

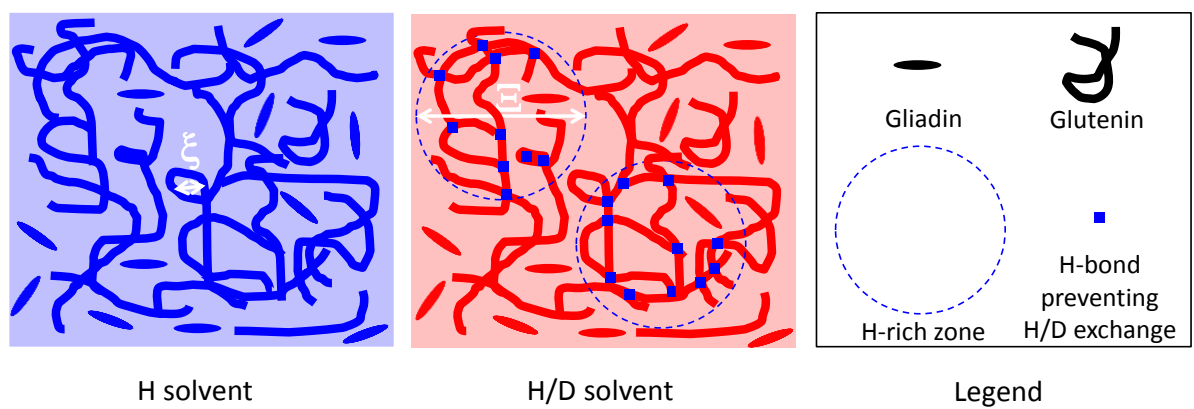

**Figure 12.** Sketch of a fractal polymeric gel of gluten. In a protonated solvent, a contrast is established between protein chains (that are characterized by a blob size $\xi$) and the solvent: SAXS and SANS

profiles are similar. In the partially deuterated solvent H-rich zones, of typical size $\Xi$, are only detectable by SANS and attributed to H-bonds that prevent H/D exchange between the solvent and some gluten polypeptides, especially glutenins.

To summarize our study, sketches of the sample structure in hydrogenated and deuterated solvents, are proposed in Figure 12. In a protonated solvent, gluten proteins form a fractal polymeric gel while a micro-phase separation is measured in a deuterated solvent. This phase separation would be attributed to the tight hydrogen bonds that exist between the glutenin polypeptides and prevent H/D exchanges.

# 5. Conclusion

We have investigated the structure of a model gluten gel prepared in ethanol/water using SANS contrast variation. A classical contrast variation analysis fails as previously observed for hydrophilic polymer solutions like PEO[12] and polyacrylamide[23] solutions. However, a careful investigation of data evidences the formation of H-rich zones for samples prepared with deuterated solvents. The analysis of dilute samples indicates that these zones are also enriched in glutenin polymers. The formation of these zones would be hence mediated by the heterogeneity of interactions within the protein components of gluten especially via hydrogen bonds. The temperature dependence of these phenomena will be studied in the next future. In addition, because hydrogen bonds are an important feature of gluten rheology as suggested by experiments performed with urea[53,22] and with deuterated water[40, 41, 54], it would be interesting to correlate the large scale structures identified here with rheological measurements.

## Acknowledgments

We thank J. Bonicel (IATE) for help in the SE-HPLC and T. Phou (L2C) for the FTIR measurements under inert atmosphere. Discussions with A.-C. Genix and J. Oberdisse are acknowledged. We would like to thank S. Perez for the assistance during the SAXS experiments. This work was supported by the Laboratoire of Excellence NUMEV (ANR-10-LAB-20). Besides, this research project has been supported by the European Commission under the 7th Framework Program through the "Research Infrastructures" action of the Capacities Program, NMI3-II, Grant Agreement number 28388 to perform the neutron scattering measurements at the Heinz Maier-Leibnitz Zentrum (MLZ), Garching, Germany.

# Graphical abstract

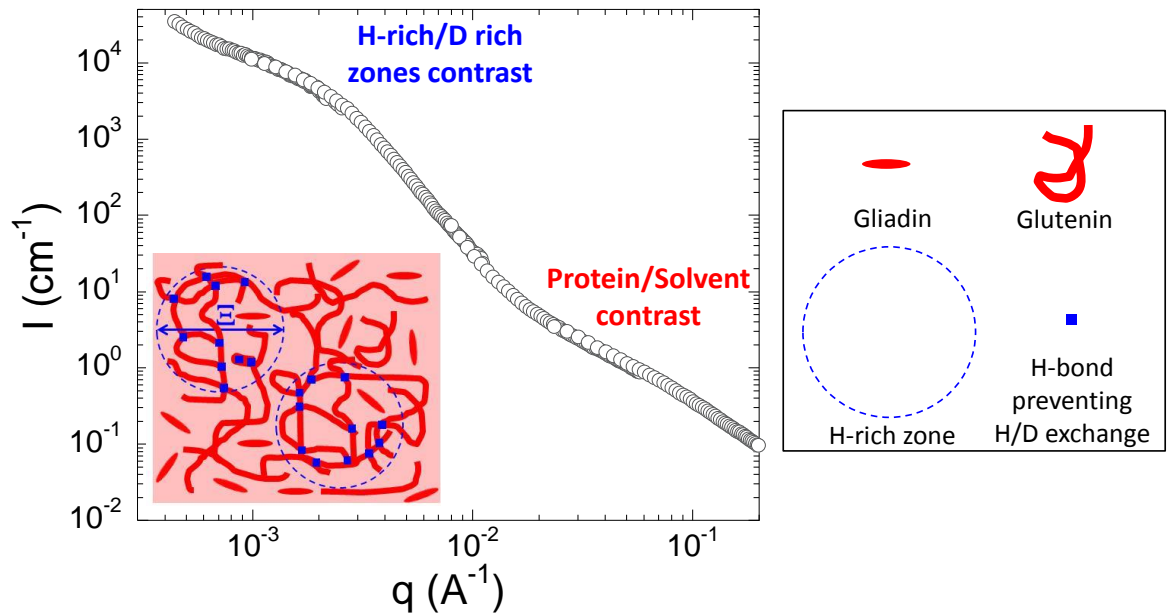

The SANS analysis of gluten gels prepared with deuterated solvent evidences the formation of large scale zones enriched in protonated proteins. The formation of these zones is associated to the heterogeneities of interaction between the different classes of gluten proteins and the solvent.

# Small angle neutron scattering contrast variation reveals heterogeneities of interactions in protein gels

*SUPPORTING INFORMATION*

A. Banc[1], C. Charbonneau[1], M. Dahesh[1,2], M-S Appavou[3], Z. Fu[3], M-H. Morel[2], L. Ramos[1]

[1] *Laboratoire Charles Coulomb (L2C), UMR 5221 CNRS-Université de Montpellier, F-34095 Montpellier, France*

[2] *UMR IATE, UM-CIRAD-INRA-SupAgro, 2 pl Pierre Viala, 34070 Montpellier, France.*

[3] *Jülich Centre for Neutron Science JCNS, Forschungszentrum Jülich, Outstation at MLZ, D-85747 Garching, Germany*

## 1. Protein extract composition

The protein composition of the gluten protein extract was assessed by size exclusion high performance liquid chromatography (SE-HPLC) and reduced SDS-PAGE analysis. The respective proportions in glutenin polymer, ω−gliadin, γ−gliadin, α/β-gliadins, and chloroform/methanol soluble (CM) proteins (which are essentially albumin and globulin, alb/glo) were estimated from the differential integration of the SE-HPLC profile of the protein extract according to Morel et al[1] (see figure S1).

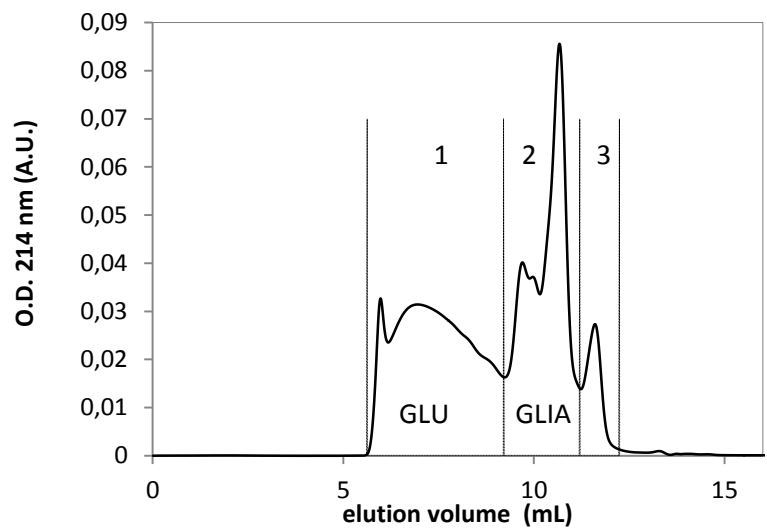

**Figure S1.** SE-HPLC profile of the wheat gluten protein fraction. The protein was dispersed in a 1% sodium-dodecyl-sulfate phosphate buffer, 20µL of the dispersion was injected on a TSK gel 4000SWXL (30 cm x 7.8 mm, 450 Å) and eluted at 0.7 ml.min$^{-1}$. The first fraction (GLU) contains glutenin polymers (100 000 < Mw < 2.10$^6$ g/mol), the second fraction contains gliadins (GLIA; 25 000 < Mw < 100 000 g/mol) and the third fraction contains small Mw proteins (<25 000 g/mol). Fractions 1, 2 and 3 account respectively for 49, 43 and 8% of total protein.

The composition of the glutenin polymer in its *x* and *y* high-molecular-weight glutenin subunits types (HMW-GS), and their proportion in total protein were obtained from the densitometric analysis of the reduced protein SDS-PAGE pattern as shown in figure S2.

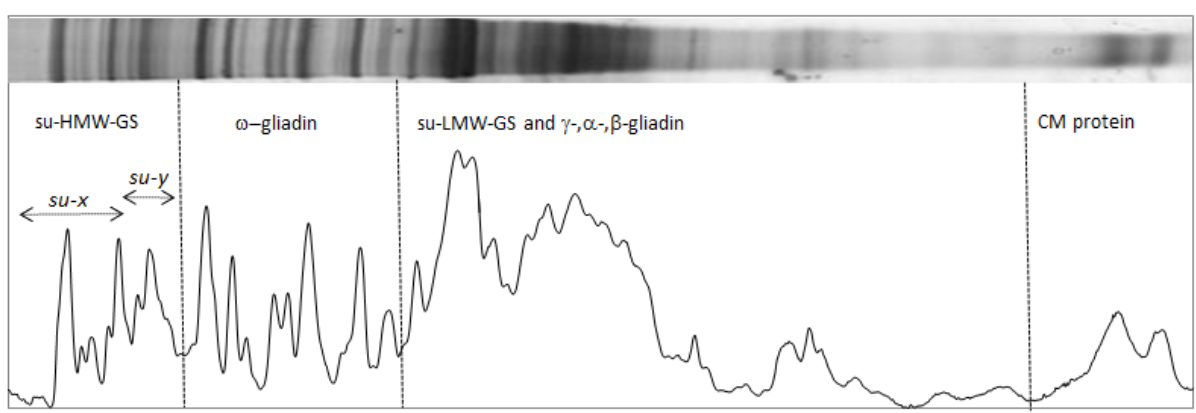

**Figure S2.** Densitometric profile of the SDS-PAGE pattern of the wheat gluten protein fraction.. Proteins were reduced with 10 mM dithioerhytritol and fractionated on a 12% SDS-PAGE prepared according to Laemmli's standard protocol. From the top (left) to the bottom of the gel: high-molecular-weight glutenin subunits of *x* and *y* types, ω–gliadins, mixture of γ–, α/β–gliadins and low-molecular-weight glutenin subunits. The last doublet consists in chloroform/methanol soluble (CM) proteins belonging to the class of α-amylase/trypsin inhibitors[2].

The proportion in low-molecular weight glutenin subunits (LMW-GS) in glutenin polymers was deduced by difference from the known proportions of high-molecular weight glutenin subunit (HMW-GS) (from SDS-PAGE) and glutenin polymers (from SE-HPLC). Similarly gliadins were distinguished into ω-gliadin and γ-gliadin or α/β-gliadin taking into consideration the results of SE-HPLC and SDS-PAGE analyses. The resulting composition of the gluten protein extract is given in table 1.

| Composition of the wheat gluten protein extract | | | | | |
|---|---|---|---|---|---|
| glutenin polymers | | | gliadin | | alb/glo |
| HMW-GSx | HMW-GSy | LMW-GS | ω-gliadin | γ, α/β gliadin | CM protein |
| 7% | 6% | 36% | 20% | 23% | 8% |

**Table 1.** Composition (in percent of total protein) of the wheat gluten protein extract as deduced from SE-HPLC and SDS-PAGE analyses.

## 2. Calculation of the average SLD of the protein extract

For the calculation of the average scattering length density (SLD) of the gluten protein extract, mean SLD values of the different wheat protein classes, namely HMW-GS type *x* and *y*, LMW-GS, ω–, γ–, α/β– gliadins and CM protein, were considered since industrial gluten is commonly obtained from a blend of different cultivars. The Jacrot[3] protonated amino-acid SLD values were used to calculate mean SLD from the known amino-acid composition of typical wheat protein. Table II presents these mean SLD values and their standard deviations calculated considering at least three representative proteins of each class.

| | SLD of wheat protein classes ($10^{-6}$ Å$^{-2}$) | | | | | | |
|---|---|---|---|---|---|---|---|
| | glutenin polymers | | | gliadin | | | alb/glo |
| | HMW-GSx[a] | HMW-GSy[b] | LMW-GS[c] | ω-gliadin[d] | γ-gliadin[e] | α/β-gliadin[f] | CM protein[g] |
| **Protonated** | 2.18 (0.01) | 2.16 (0.01) | 1.93 (0.04) | 2.05 (0.05) | 1.98 (0.04) | 1.98 (0.02) | 1.84 (0.03) |
| **Deuterated** | 3.75 (0.04) | 3.73 (0.04) | 3.32 (0.07) | 3.35 (0.13) | 3.27 (0.07) | 3.35 (0.03) | 3.2 (0.2) |

The following UniProtKB accession were considered for calculation. Standard deviation in brackets.

[a] P10388, P08489, Q1KL95, Q599I0, Q6UKZ5, H9B854, Q0Q5D8.

[b] P08488, Q0Q5D8, A9ZMG8.

[c] Q8W3V2, P10386, P16315, Q8W3V5, Q00M61, Q6SPZ1, Q5MFQ2, Q6SPY7, B2BZD1, B2Y2R3, Q8W3X2.

[d] C0KEI0, Q571R2, R9XWH8, A0A060N0S6, C0KEI1, C0KEH9, A0A0B5J8A9, A0A0B5JD20, A0A0B5JHW1.

[e] P08079, P08453, P06659, P21292, P04729, P04730, M9TK56, R9XUS6.

[f] P18573, P04724, P04723, P02863, P04721, P04722, P04725, H6VLP5, A5JSA4.

[g] P93594, P16159, A8R0D1, A9JPD1, P30110.

**Table 2.** SLD of the different peptide classes identified in the wheat gluten fraction. Values for fully protonated proteins and values for proteins with 100% of exchangeable hydrogen replaced by deuterium are indicated.

The mean SLD value of the gluten protein extract calculated form the contribution of each protein classes (Table 1) and their individual SLD values (Table 2) is $(1.99 \pm 0.14)\ 10^{-6} \text{Å}^{-2}$. The standard deviation $(0.14\ 10^{-6} \text{Å}^{-2})$ takes into account the standard deviation on the SLD of each class of protein but also a 5% uncertainty on their specific contribution to the total protein content of the wheat gluten fraction. The same kind of calculation was performed considering that all exchangeable hydrogen atoms are replaced by deuterium (Figure S3). The mean SLD value shifts from $1.99\ (\pm 0.14)$ to $3.4\ (\pm 0.3)\ 10^{-6} \text{Å}^{-2}$.

## 3. Phase separation at Φ=0.04

Figure S4 displays pictures of samples prepared at Φ=0.04 with purely protonated and purely deuterated solvents.

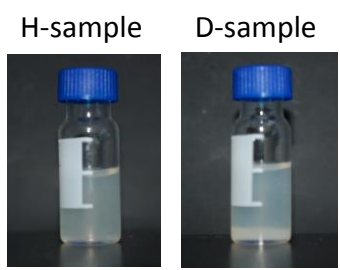

**Figure S4.** Pictures of samples prepared at Φ=0.04 with different solvents: $H_2O/C_2H_5OH$ (50/50 v/v) for the H-sample, $D_2O/C_2H_5OD$ for the D-sample.